\documentstyle[aps,prl,epsf,twocolumn]{revtex}
\begin{document}
\draft
\twocolumn[\hsize\textwidth\columnwidth\hsize\csname @twocolumnfalse\endcsname
\title{Melting of the Electron Glass}
\author{A. A. Pastor and V. Dobrosavljevi\'{c}}
\address{Department of Physics and 
National High Magnetic Field Laboratory, \\Florida State University, 
Tallahassee, Florida 32306}

\date{\today}
\maketitle

\begin{abstract}

A model of spinless interacting electrons in presence of disorder is
examined using an extended dynamical mean-field formulation.  When the
interaction strength is large as compared to the Fermi energy, a low
temperature glassy phase is identified, which in our formulation
corresponds to a replica-symmetry breaking instability. The glassy
phase is characterized by a pseudo-gap in the single particle density
of states, reminiscent of the Coulomb gap of Efros and Shklovskii.
Due to ergodicity breaking, the ``zero-field cooled'' compressibility
of this electron glass vanishes at T=0, consistent with absence of
screening.  When the Fermi energy exceeds a critical value, the glassy
phase is suppressed, and normal metallic behavior is recovered.

\end{abstract}

\pacs{PACS Numbers: 75.20.Hr, 71.55.Jv}  
]
\narrowtext

\mbox{}

Understanding the basic physical processes determining whether a
material is a conductor or an insulator continues to be a central
theme of Condensed Matter Physics.  In recent years, much progress has
been achieved in understanding weakly disordered metals \cite{lr}, but
the physics of the metal-insulator transition (MIT) in presence of
disorder remains largely an open problem.

The interplay of the electron-electron interactions and disorder is
particularly evident deep on the insulating side of the MIT.  Here,
both experimental \cite{marklee,shklovskii} and theoretical studies
\cite{efros} have demonstrated that they can lead to the formation of
a soft ``Coulomb gap'', a phenomenon that is believed to be related to
the glassy behavior of the electrons. Such glassy freezing has long
been suspected \cite{tedglass} to be of importance, but very recent
work \cite{sudip} has suggested that it may even dominate the
MIT behavior in certain low carrier density systems.  The classic work
of Efros and Shklovskii \cite{efros} has clarified some basic aspects
of this behavior, but a number of key questions remain.  In
particular: (1) What is the precise relation of this glassy behavior
and the emergence of the Coulomb gap? (2) What should be the order
parameter for the glass phase? (3) How should the glassy freezing
affect the compressibility and the screening of the electron gas? (4)
How do the quantum fluctuations (electron tunneling) melt this glass
and influence the approach to the MIT?

In this Letter, we examine a simple model for which all these
questions can be rigorously answered in the limit of large
coordination.  Within our model, the universal form of the Coulomb gap
proves to be a direct consequence of glassy freezing. The glass phase
is identified through the emergence of an extensive number of
metastable states, which in our formulation is manifested as a replica
symmetry breaking instability. As a consequence of this ergodicity
breaking, the zero-field cooled compressibility is found to vanish at
T=0, suggesting the absence of screening in disordered
insulators. Finally, we show that quantum fluctuations can melt this
glass even at $T=0$, but that the relevant energy scale is set by the
electronic mobility, and is therefore a nontrivial function of
disorder.

In order to concentrate on the physics of the electron glass, we focus
on a simple lattice model of spinless interacting electrons in
presence of randomness, as given by the Hamiltonian

\begin{equation}
H=\sum_{ij} ( -t_{ij} + \varepsilon_i \delta_{ij})
c^{\dagger}_{i}c_{j}\nonumber +
\sum_{ij} V_{ij} n_i n_j.
\end{equation}

Here, $n_i = c^{\dagger}_{i} c_i$ are the occupation number operators,
$t_{ij}$ are the corresponding hopping elements, $V_{ij}$ describes
the inter-site electronic repulsion, and $\varepsilon_i$ are the
random site energies.  Motivated by the long range nature of the
Coulomb interaction, we wish to examine the situation where a given
electron interacts with a large number $z_v$ of other electrons. The
analysis is simplest if the interaction $V_{ij}=V$ is taken to be
uniform within a volume containing $z_v$ neighbors, as opposed to the
more realistic Coulomb interactions. Nevertheless, most of the {\em
qualitative} features of the Coulomb glass are still captured if we
consider $z_v >> 1$, as follows.

To obtain the relevant dynamical mean-field (DMF) equations
\cite{dmf}, we focus on a given lattice site, and formally integrate
out \cite{dmf} all the other degrees of freedom.  For large
coordination, the corresponding local effective action simplifies,
since it suffices to consider contributions up to the second order in
the hopping elements $t$ and the interaction amplitudes
$V$\cite{dinflimit}. We concentrate on a Bethe lattice at half
filling, where the equations are simplest, and we find

\begin{eqnarray}
&&S_{eff} (i)
=\sum_{a}\int_o^{\beta}\int_o^{\beta }d\tau d\tau '\;[
c^{\dagger a}_i (\tau )( \delta (\tau -\tau ')
\partial_{\tau} +\varepsilon_i \nonumber \\
&&+ t^2 G (\tau,\tau ') )c^{a}_i (\tau ')
+\frac{1}{2}V^2\delta n_i^{a}(\tau )\chi (\tau ,\tau ')
\delta n_i^{a} (\tau ')]\nonumber\\
&&+\frac{1}{2}V^2\sum_{a\neq b}\int_o^{\beta }\int_o^{\beta }d\tau d\tau '
\; \delta n_i^{a}(\tau )\; q_{ab}\; 
\delta n_i^{b} (\tau ').
\end{eqnarray}

We have used standard functional integration over replicated Grassmann
fields \cite{dinfdis}, where $a=1,...n$ ($n\rightarrow 0$) are the 
replica indices\cite{replica}.  Here, the operators $\delta
n_{i}^{a}(\tau )=( c^{\dagger a}_{i} (\tau )c_{i}^{a} (\tau ) - 1/2)$
represent the {\em density fluctuations} from half filling. The order
parameters $G (\tau -\tau ')$, $\chi(\tau -\tau ')$ and $q_{ab}$
satisfy the following set of self-consistency conditions

\begin{eqnarray}
&&G (\tau -\tau ')=\int d\varepsilon_i P(\varepsilon_i )
< c^{\dagger a}_{i} (\tau )c_{i}^{a} (\tau ')>_{eff},\\
&&\chi (\tau -\tau ')=\int d\varepsilon_i P(\varepsilon_i )
< \delta n^{\dagger a}_{i} (\tau )\delta n_{i}^{a} (\tau ')>_{eff},\\
&&q_{ab}=\int d\varepsilon_i P(\varepsilon_i )
< \delta n^{\dagger a}_{i} (\tau )\delta n_{i}^{b} (\tau ')>_{eff}.
\end{eqnarray}

In these equations, the averages are taken with respect to the
effective action of Eq. (2), and $P(\varepsilon_i )$ is the
probability distribution of random site energies.  Physically, the
quantity $G (\tau -\tau ')$ represents the single-particle electronic
spectrum of the environment, as seen by an electron on site $i$.  The
second quantity $\chi(\tau -\tau ')$ is the averaged dynamic
compressibility.  This {\em mode-coupling} term reflects the retarded
response of the density fluctuations of the environment.  Finally, the
quantity $q_{ab}$ $(a\neq b)$ is the familiar Edwards-Anderson order
parameter \cite{parisi}. Its nonzero value indicates that the time
averaged electronic density is spatially non-uniform.

We should emphasize that in presence of randomness, the electronic
density will remain {\em nonuniform} at any temperature. As a result,
the Edwards-Anderson order parameter $q = \overline{<\delta n>^2}\neq
0$ everywhere, and thus cannot be used to identify the glass
transition. However, in the classical ($t=0$) limit, a simple
transformation $S_i = 2\delta n_i$, $J=V/4$, and $h_i = \varepsilon_i
/2$ reduces our problem to the familiar random-field Ising
model. Here, recent work \cite {mezard} has suggested that a glassy
phase can be identified at low temperatures by carrying out an
appropriate replica symmetry breaking (RSB) stability analysis.

To carry such a RSB analysis in our approach, note that the our
self-consistency conditions Eqs. (2)-(5) can be derived from an
equivalent {\em infinite range} model given by Eq. (1), but having
{\em random} hopping elements $t_{ij}$ and {\em random} inter-site
interactions $V_{ij}$ \cite{florian}, with zero mean and variance $t^2
/N$, and $V^2 /N$, respectively ($N$ is the number of lattice
sites). This model is similar to familiar spin-glass models, and the
RSB analysis can be carried out using standard methods \cite{parisi}.
The resulting stability criterion takes the form

\begin{equation}
1 = V^2 \int d\varepsilon_i P(\varepsilon_i )
\left[\chi_{loc} (\varepsilon_i )\right]^2.
\end{equation}
Here, $\chi_{loc} (\varepsilon_i )$ is the {\em unaveraged} local
compressibility, that can be expressed as
\begin{equation}
\chi_{loc} (\varepsilon_i )= 
\frac{\partial}{\partial\varepsilon_i }
\frac{1}{\beta}\int_{o}^{\beta}d\tau
<\delta n(\tau )>_{S_{eff} (i)}.
\end{equation}

We continue by concentrating on the classical $(t=0)$ limit, where
$\chi (\tau ,\tau ' ) =1/4$, and the problem can easily be solved in
closed form. We first focus on the replica symmetric (RS) solution,
which is valid at high temperatures, and set $q_{ab}=q$ for all
replica pairs, giving
\begin{equation}
q = \frac{1}{4}\int_{-\infty}^{+\infty}\frac{dx}{\sqrt{\pi}}
e^{-x^2/2}\tanh^2\left[ \frac{1}{2}x\beta W_{eff} (q)\right] ,
\end{equation}
where we have introduced an effective disorder strength
$W_{eff}=\sqrt{W^2 +V^2q}$ , and have considered a Gaussian
distribution of random site energies of variance $W^2$. Even for weak
disorder, $q\neq 0$ at any temperature, but it becomes appreciable
only for $T\le V/4$. These frozen-in density fluctuations introduce an
added component to the random potential seen by the electron.  In this
way $W_{eff}$ can become appreciable at low temperatures even if the
``bare'' disorder $W$ is weak; this represents an interaction-driven
mechanism for resistivity enhancement.

Next, we examine the stability of this RS solution. From Eq.  (6) we
find an expression for the instability line, which takes the form
\begin{equation}
1=\frac{(\beta V)^2}{16}
\int_{-\infty}^{+\infty}\frac{dx}{\sqrt{\pi}}
e^{-x^2/2}\cosh^{-4}\left[ \frac{1}{2}x
\beta W_{eff}(q)\right],
\end{equation}
where $q(T,W)$ is given by Eq. (8). The resulting expression for the
glass transition temperature simplifies for $W >> V$, where using the
fact that $q\le 1/4$ we find $T_G \approx
\frac{1}{6\sqrt{2\pi}}\frac{V^2}{W}$. This reduction of $T_G$ at large
$W$ reflects the {\em reduction} of frustration in the strong disorder
limit, where the electrons simply populate the lowest potential
wells. Still, in contrast to the well known ``AT-line'' \cite{parisi},
$T_G$ decreases slowly with disorder, suggesting that the glassy
behavior of electrons may be observable at finite temperatures,
in agreement with some recent experiments \cite{goldman}.

To understand this behavior, we investigate the structure of the
low-temperature glass phase. Consider the single-particle density of
states at T=0, which in the classical limit can be expressed as
\begin{equation}
\overline{\rho} (\varepsilon , t=0)= \frac{1}{N}
\sum_i \delta (\varepsilon -\varepsilon^{R}_i ), 
\end{equation}
where $\varepsilon_i^{R} \equiv \varepsilon_i + \sum V_{ij} n_j$ are
the renormalized site energies.  In the thermodynamic limit, this
quantity is nothing but the probability distribution
$P_R (\varepsilon^{R}_i )$.  It is analogous to the ``local field
distribution'' in the spin-glass models, and can be easily shown to
reduce to a simple Gaussian distribution in the RS theory,
establishing the {\em absence} of any gap for $T>T_G$. Obtaining
explicit results from a replica calculation in the glass phase is more
difficult, but useful insight can be achieved by using standard
simulation methods \cite{palmer,zimanyi} on our equivalent
infinite-range model; some typical results are shown in Fig. 1.
\begin{figure}
\epsfxsize=3.2in \epsfbox{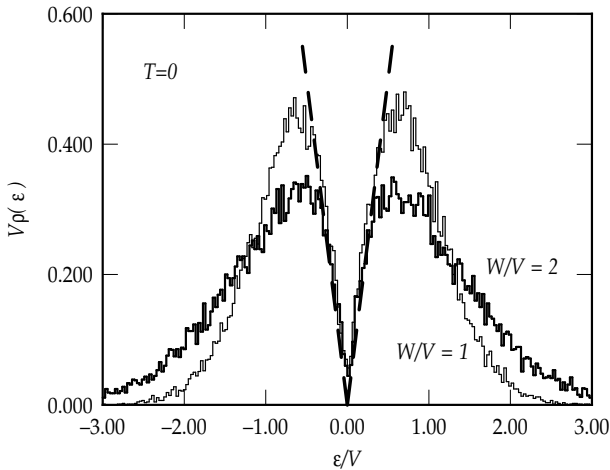}
\caption{Single particle density of states in the 
classical ($t=0$) limit at $T=0$, 
as a function of disorder strength. Results are shown 
from a simulation on $N=200$ site system, 
for $W/V = 0.5$ (thin line) and $W/V = 1.0$ (full  line). 
Note that the low energy form of the gap takes a {\em universal} form,
independent of the disorder strength $W$. The dashed line follows 
Eq. (11).}
\end{figure}
We find that as a result of glassy freezing, a pseudo-gap emerges in
the single-particle density of states, reminiscent of the Coulomb gap
of Efros and Shklovskii (ES) \cite{efros}. The low energy form of this
gap appears {\em universal},
\begin{equation}
\rho (\varepsilon)  \approx C|\varepsilon |^{\alpha}/V^2;
\;\;\; C=\alpha = 1,
\end{equation}
independent of the disorder strength $W$, again in striking analogy
with the predictions of ES\cite{esexponent}. To establish this result,
we have used stability arguments very similar to those developed for
spin-glass (SG) models \cite{zimanyi}, demonstrating that the form of
Eq. (11) represents an exact {\em upper bound} for $\rho(\varepsilon
)$. For {\em infinite-ranged} SG models, as in our case, this bound
appears to be {\em saturated}, leading to universal behavior. Such
universality is often associated with a critical, self-organized state
of the system. Recent work \cite{zimanyi} finds strong numerical
evidence of such criticality for SG models; we believe that the
universal gap form in our case has the same origin. Furthermore,
assuming that the universal form of Eq. (11) is obeyed immediately
allows for an estimate of $T_G (W)$.  Using Eq. (11) to estimate the
gap size for large disorder gives $T_G \sim E_g \sim V^2 /W$, in
agreement with Eq. (9).

The ergodicity breaking associated with the glassy freezing has
important consequences for our model. Again, using the close
similarity of our classical infinite range model to standard SG models
\cite{parisi}, it is not difficult to see that the {\em zero-field
cooled} (ZFC) compressibility vanishes at $T=0$, in contrast to the
field-cooled one, which remains finite.  Essentially, if the chemical
potential is modified {\em after} the system is cooled to $T=0$, the
system immediately falls out of equilibrium and displays hysteretic
behavior \cite{zimanyi} with vanishing {\em typical}
compressibility. If this behavior persists in finite dimensions and
for realistic Coulomb interactions, it could explain the absence
of screening in disordered insulators.  We emphasize that our picture
does not agree with the recently proposed scenario of Si and Varma
\cite{varma}, where the two compressibilities are not distinguished,
and are {\em both} proposed to vanish in the $T=0$ insulating state.

Finally, we examine the effect of quantum fluctuations on the
stability of the glass phase. We generally expect that for a
sufficiently large Fermi energy $E_F\sim t_G$, $T_G\rightarrow 0$, but
of particular interest is the disorder dependence of $t_G (W)$. In the
quantum case, solving our DMF equations becomes a non-trivial
many-body problem, which cannot be solved in closed form. The problem
simplifies in the strong disorder limit, where the exact expression
for the RSB boundary can be obtained. As in the classical case, to
examine the $W >> V$ limit, we use the fact that $q \leq 1/4$, $\chi
(\tau -\tau ')\leq 1/4$, so that in solving Eq. (6) we can eliminate
the corresponding terms in the expression for the local effective
action Eq. (2).  The calculation of the local compressibilities of
Eq. (3) then reduces to computing the same quantity for noninteracting
electrons in presence of disorder, and we find
\begin{equation}
\chi_{loc} (\varepsilon )
=\frac{\beta}{4}\int_{-\infty}^{+\infty} d\omega
\rho^{\infty }_{\varepsilon}(\omega )
\cosh^{-2} (\frac{1}{2}\beta\omega ).
\end{equation}
Here, the {\em local} density of states (LDOS) reduces to a narrow
resonance of width $\Delta =\pi t^2 P(\varepsilon_i =0)\sim t^2/W$
\begin{equation}
\rho^{\infty }_{\varepsilon}(\omega )
\approx \frac{1}{\pi } \frac{\Delta }{(\omega -\varepsilon )^2 
+\Delta^2},
\end{equation}
and we get
\begin{equation}
t_G (T=0,
W\rightarrow\infty )= V/\pi\sqrt{2}.
\end{equation}
In contrast to the effect of finite temperatures, we find that
a {\em finite} value of the Fermi energy $E_F \sim t$ is
needed to suppress the glass phase, even for $W >> V$ ! This
surprising result has in fact a simple physical interpretation. At 
$t=0$ we have argued that the glass phase for $W >> V$ is 
characterized by a small energy scale $E_g \sim V^2 /W$,
hence the reduction of $T_G$ at strong disorder. In the 
$T=0$ quantum case, the role of the thermal energy is 
replaced by an energy scale measuring the size of
quantum fluctuations. However, the relevant energy scale is
{\em not} the Fermi energy $E_F \sim t$, but instead is given
by the {\em hybridization energy} $\Delta \sim t^2 /W$
which, according to Fermi's golden rule, determines
the inverse lifetime of an electron on a given site. 
Due to the decrease of the  electronic mobility at large disorder, 
$\Delta (W)$ is reduced at $W >> t$; comparing it
to the $T=0$ gap size $E_g (W)$ immediately explains Eq. (14).

Our solution, which evaluates the local compressibilities in Eq. (6)
by setting $V=0$ gives the exact behavior for $W >> V$.  This
approximation can be considered as a leading term in a
``weak-coupling'' approach, where the compressibilities are expanded
in powers of the interaction, similar in spirit to the RPA of the
electron gas. For our purposes of calculating the phase diagram, this
strategy will provide qualitatively correct results even for weaker
disorder, and the resulting phase diagram is shown in Fig. 2.  Note
that our glass phase persists even as $W\rightarrow 0$. However, in
this regime we expect uniform charge ordering to preempt any glass
formation, an effect that we have not examined, but that can easily be
incorporated in our framework.\vspace{-12pt}
\begin{figure}
\epsfxsize=3.2in \epsfbox{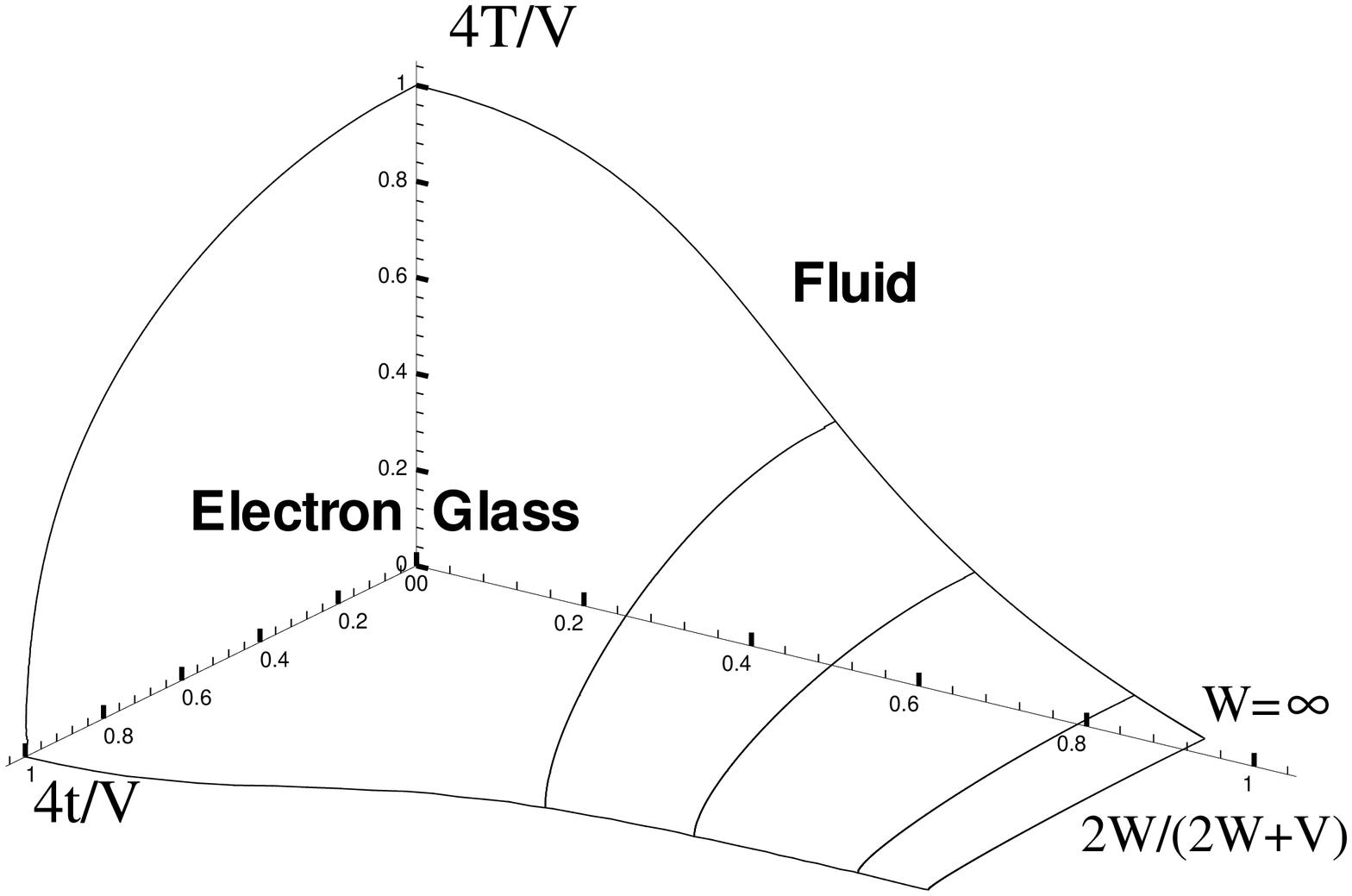}
\caption{Phase diagram as a function of temperature T, disorder
strength W, and hopping element $t$. Note how the glass
transition temperature $T_G$ decreases  as $~1/W$ in the
strong disorder limit. In contrast, the critical value of the hopping 
element $t_G$ remains {\em finite} as $W\rightarrow\infty$.}
\end{figure}
Our results clearly demonstrate that the quantum fluctuations become a
nontrivial function of disorder in this electronic context, which
plays a key role in controlling the glassy phase.  We should emphasize
that, within our theory, the relevant hybridization energy $\Delta$
remains {\em finite} at any $t\neq 0$, since Anderson localization
\cite{anderson} is absent at $z_t =\infty$. As a result,
while the glass phase will persist for all $t < t_G$, the systems will
become metallic for any $t\neq 0$.  Similarly as finite temperatures,
$\Delta\neq 0$ will introduce a cutoff for both our Coulomb gap,
and the corresponding ZFC compressibility.  On the other hand,
Anderson has demonstrated \cite{anderson} that $\Delta \rightarrow 0$
in an Anderson insulator, even in absence of interactions.  We thus
expect that Anderson localization will strongly stabilize the glass
phase, as well as restore the ZFC incompressibility and the existence
of the Coulomb gap in the entire insulating phase. Our theory can be
readily extended to incorporate the Anderson localization effects
within the DMF framework, following the approaches of
Ref. \cite{motand}, but these interesting issues will be addressed
elsewhere.

To conclude, we have presented a simple theoretical framework where
several aspects of glassy freezing of electrons can be investigated in
detail. We find that glassy freezing strongly enhances the
effective disorder seen by the electrons, providing a new driving
force for electron localization. In addition, the ergodicity breaking
in the glassy state leads to the universality of the Coulomb gap, as
well as the incompressible behavior in disordered insulators.

We thank E. Abrahams, A. Georges, M. Horbach, T. R. Kirkpatrick,
D. Popovi\'{c}, A. E. Ruckenstein, Q. Si, G. Zarand, and G. Zimanyi
for useful discussions.  This work has been supported in part by the
NSF grant DMR-9974311, the NHMFL/FSU, and the Alfred P. Sloan
Foundation.
\vspace{-12pt}


\begin{references}
\bibitem{replica} For mean-field glassy models \cite{parisi}, such as
ours, there exists considerable evidence that most results obtained
using replicas can also be derived using alternative approaches.
\bibitem{esexponent} The result $\alpha =1$ that we find is specific
to our model. In early work, Efros and Shklovskii \cite{efros} have
suggested that the precise value of this exponent should depend on
both the (power-law) form of the interaction and the dimensionality,
but should display universality by being insensitive to the disorder
strength - in agreement with our findings.
\end{references}
\end{document}